# True decoherence-free-subspace derived from a semiconductor double quantum dot Heisenberg spin-trimer

Wonjin Jang[1,2], Jehyun Kim[1,3], Jaemin Park[1], Min-Kyun Cho[1,4], Hyeongyu Jang[1], Sangwoo Sim[1], Hwanchul Jung[5], Vladimir Umansky[3], and Dohun Kim[1]*

[1]NextQuantum, Department of Physics and Astronomy, and Institute of Applied Physics, Seoul National University, Seoul 08826, Korea

[2]Institute of Physics and Center for Quantum Science and Engineering, École polytechnique fédérale de Lausanne (EPFL), Lausanne, 1015, Switzerland

[3]Braun Center for Submicron Research, Department of Condensed Matter Physics, Weizmann Institute of Science, Rehovot 76100, Israel

[4]Center for Superconducting Quantum Computing System, Korea Research Institute of Standards and Science, Daejeon 34113, Korea

[5]Department of Physics, Pusan National University, Busan 46241, Korea

*Corresponding author: dohunkim@snu.ac.kr

**Abstract**

Spins in solid systems can inherently serve as qubits for quantum simulation or quantum information processing. Spin qubits are usually prone to environmental magnetic field fluctuations; however, a spin qubit encoded in a decoherence-free-subspace (DFS) can be protected from certain degrees of environmental noise depending on the specific structure of the DFS. Here, we derive the "true" DFS from an antiferromagnetic Heisenberg spin-1/2 trimer, which protects the qubit states against both short- and long-wavelength magnetic field fluctuations. We define the spin trimer with three electrons confined in a gate-defined GaAs double quantum dot (DQD) where we exploit Wigner-molecularization in one of the quantum dots. We first utilize the trimer for dynamic nuclear polarization (DNP), which generates a sizable magnetic field difference, $\Delta B_z$, within the DQD. We show that large $\Delta B_z$ significantly alters the eigenspectrum of the trimer and results in the "true" DFS in the DQD. Real-time Bayesian estimation of the DFS energy gap explicitly demonstrates protection of the DFS against short-wavelength magnetic field fluctuations in addition to long-wavelength ones. Our findings pave the way toward compact DFS structures for exchange-coupled quantum dot spin chains, the internal structure of which can be coherently controlled completely decoupled from environmental magnetic fields.

The semiconductor quantum dot (QD) system represents a platform for the exploration of interacting electrons where the electrons confined in QDs are often described by the Fermi-Hubbard model[1,2]. In particular, when direct charge transitions between adjacent QDs are forbidden, in analogy to the Mott-insulating phase[3], QDs can still exchange their spins via exchange interactions[4,5]. These systems allow quantum simulations[6–8] and magnetic-field-free encoding of spin qubits[9,10], and can thus be considered a scalable resource for practical quantum computations.

While the environment coupled to a qubit can lead to undesired decoherence, qubit states encoded in a decoherence-free-subspace (DFS) can be protected against certain fluctuations in the environment[11,12]. Several types of QD qubit encoding in DFS have been proposed considering protection from electric and long- and short-wavelength magnetic noise compared with the system size[13]. Examples include charge quadruple DFS to overcome dipolar charge noise[14] or singlet-triplet qubits to protect against long-wavelength magnetic field fluctuations[4,15]. Recently, more elaborate schemes such as singlet-only encoding in a triple QD (TQD)[16] have been theoretically shown to form a "true" DFS, which screens both

short- and long-wavelength fluctuations. However, explicit demonstrations of such yet remain elusive, where a simpler structure may facilitate experimental realization of the "true" DFS.

The derivation of a DFS requires tunable energy levels which are accessible within the experimental bandwidth[14,16,17], where Coulomb correlation effects may provide controllable low-lying energy states for such applications[18,19]. Initially proposed as a few-electron counterpart of a Wigner crystal[20], Coulomb correlation-driven Wigner molecularization in QDs[18,19,21,22] has been drawing much attention for its relevance to quantum information science[19,23,24]. Along with ground state localization[25], the Wigner molecularization further quenches the orbital splitting from $10^2$ $\mu$eV to the microwave-accessible $10^0$ $\mu$eV regime, as demonstrated in carbon-nanotube[26], Si[27], GaAs,[23,24] and Ge[28] QDs. In addition to spin qubit operations exploiting the reduced energy gap of the Wigner molecule (WM) states[23], engineering of the states may further promote coherent DFS structures, which stays unexplored to date.

In this work, we derive a "true" DFS robust to both short- and long-wavelength magnetic fluctuation from an exchange-coupled spin-1/2 trimer. We define the trimer with three electrons confined in a gate-defined GaAs double QD (DQD), where the Wigner-molecularization in one of the QDs quenches the excitation spectrum of the trimer down to < 2 GHz·$h$ ($h$ is Planck's constant). Exploiting the dynamic nuclear polarization (DNP) enabled by the Wigner-molecularization[24], we build the spatial magnetic field difference, $\Delta B_z$, between the two QDs. A large $\Delta B_z$ is shown to significantly alter the eigenstates of the trimer and result in a "true" DFS in the DQD, in contrast to previously proposed schemes requiring a TQD[16] or a quadruple QD array[17]. Real-time Bayesian Hamiltonian estimation[29–31] confirms that the fluctuation of the energy gap between DFS states is smaller than that of the states outside the DFS, proposing a magnetic-noise-resilient spin qubit based on exchange interactions in a DQD.

# True decoherence-free-subspace in a Heisenberg spin-1/2 trimer

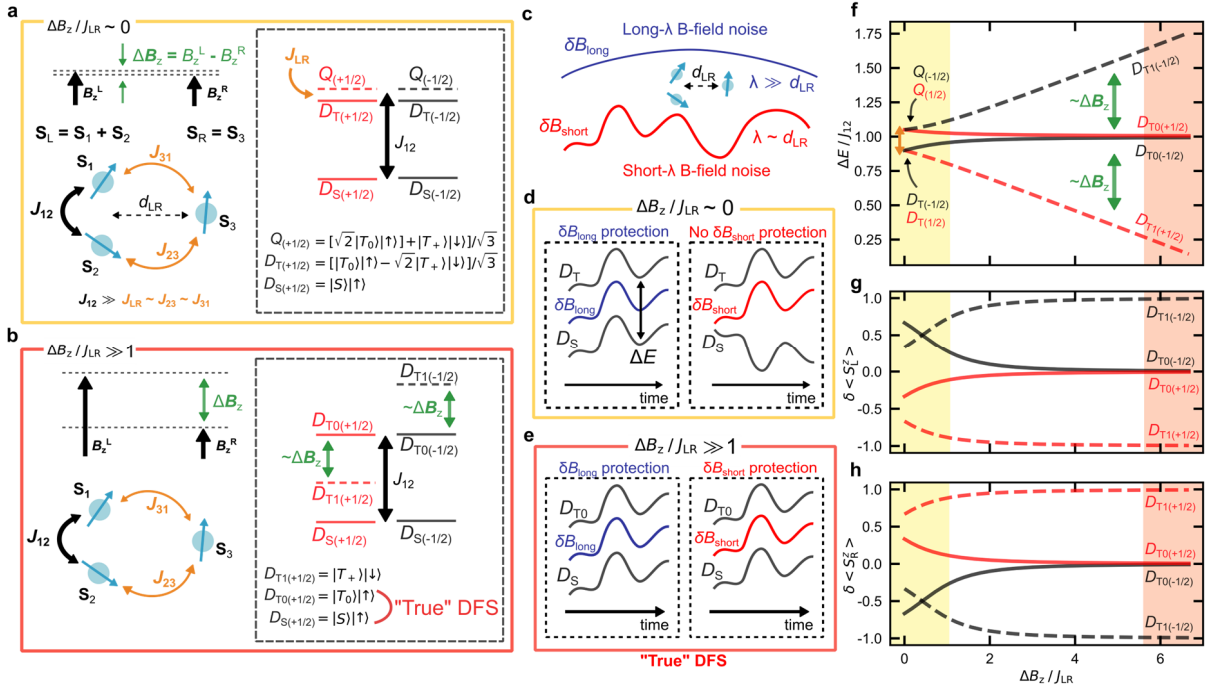

**Fig. 1 a,** Antiferromagnetic Heisenberg spin-1/2 trimer under a small magnetic field gradient. $S_1$, $S_2$, and $S_3$ denote the spin states forming the trimer, where $J_{ij}$ is the antiferromagnetic exchange interaction between $S_i$ and $S_j$. $S_1$ and $S_2$ are placed in close proximity in the site on the left, whereas $S_3$ is placed in the site on the right relatively far away from $S_1$ and $S_2$, resulting in $J_{12} \gg J_{LR} \sim J_{23} \sim J_{31}$ with $J_{LR}$ being the inter-site exchange interaction. $B_z^L$ ($B_z^R$) represents the magnetic field along the z-axis in the left (right) site. Inset (black dashed box): Eigenstates of the described trimer in $m_s = +1/2$ (red) and $m_s = -1/2$ (black) subspace under small magnetic field gradient $\Delta B_z = B_z^L - B_z^R$. $D_S$ and $D_T$ states define a qubit in each $m_s$ subspace. The $|m_s| = 3/2$ states are omitted for simplicity (see Methods for explicit expression of the spin states). **b,** Antiferromagnetic Heisenberg spin-1/2 trimer under large magnetic field gradient. Inset (black dashed box): Eigenstates of $m_s = +1/2$ (red) and $m_s = -1/2$ (black) subspace altered by large $\Delta B_z \gg J_{LR}$ (see text for the unit). $D_S$ and $D_{T0}$ states form a "true" decoherence-free-subspace robust to both long- and short-wavelength magnetic field noise. **c,** Schematic of long- (short-) wavelength magnetic field noise $\delta B_{long}$ ($\delta B_{short}$) shown as a blue (red) curve. The characteristic length scale of $\delta B_{long}$ ($\delta B_{short}$) is longer than (comparable to) the physical scale of the trimer, $d_{LR}$. **d (e),** Schematic of $D_{T(\pm 1/2)}$–$D_{S(\pm 1/2)}$ ($D_{T0(\pm 1/2)}$–$D_{S(\pm 1/2)}$) energy gap fluctuation along $\delta B_{long}$ and $\delta B_{short}$ in time. **f,** Excitation energy spectrum of the Heisenberg spin trimer as a function of $\Delta B_z / J_{LR}$. Each curve illustrates the energy gap between the denoted state and the corresponding ground state $D_{S(+1/2)}$ or $D_{S(-1/2)}$. For the calculation $J_{LR} = J_{23} = J_{13} = 0.1 J_{12}$ is assumed. The orange arrow at $\Delta B_z / J_{LR} = 0$ denotes the size of $J_{LR}$. **g (h),** Susceptibility to local magnetic field fluctuation in the left (right) site, as a function of $\Delta B_z / J_{LR}$. The difference in average z-axis spin value in the left (right) site, between the denoted state and the corresponding ground state, is evaluated (see text for details.).

We first elucidate the "true" DFS which can be found in a Heisenberg spin-1/2 trimer. Figure 1a illustrates three spin-1/2 particles interacting with each other by antiferromagnetic exchange interaction. Here, $S_i$ denotes the spin state of the i$^{th}$ particle and $J_{ij}$ represents the exchange interaction between $S_i$ and $S_j$. We assume a trimer where $S_1$ and $S_2$ are placed close to each other in the site on the left, and $S_3$ is placed in the site on the right, separated by $d_{LR}$ from that on the left. This results in $J_{12} \gg J_{LR} \sim J_{23} \sim J_{31}$, where $J_{12}$ ($J_{LR}$) presents intra (inter) site exchange interaction. Taking into account the magnetic field difference $\Delta B_z = B_z^L - B_z^R$ between the sites, the Hamiltonian of the system is described below (Eqn. 1). We consistently use a unit with $g^* \mu_B = 1$ to normalize the magnetic field to the energy ($g^*$ is the gyromagnetic ratio of the spin, and $\mu_B$ is the Bohr magneton.).

$$H = J_{12} \mathbf{S}_1 \cdot \mathbf{S}_2 + J_{23} \mathbf{S}_2 \cdot \mathbf{S}_3 + J_{31} \mathbf{S}_3 \cdot \mathbf{S}_1 + \Delta B_z/2\, (\mathbf{S}_1^z + \mathbf{S}_2^z - \mathbf{S}_3^z) \qquad \text{(Eqn. 1)}$$

At $\Delta B_z < J_{LR}$, the Hamiltonian results in doublet and quadruplet ($Q$) states with the total spin quantum number $S_{tot} = 1/2$ and $3/2$, respectively[9,24]. Depending on the symmetry of the spin states, a doublet state can form either a doublet-singlet ($D_S$) or a doublet-triplet ($D_T$). In our case with $J_{12} \gg J_{LR}$, $J_{12}$ dominates the $D_S$-$D_T$ energy gap (inset to Fig. 1a) where $D_S$ ($D_T$) involves the singlet (triplet) pairing of $S_1$ and $S_2$ in the left site (see Methods for explicit spin expressions). We note that the $D_T$-$Q$ energy gap is given by $J_{LR}$. The finite magnetic field splits each doublet (quadruplet) state into $m_s = +1/2$ and $-1/2$ ($+3/2$, $+1/2$, $-1/2$ and $-3/2$) states, where $m_s$ represents the spin projection to the z-axis. We introduce a notation $A_{(m_s)}$ to denote the spin state $A$ and the corresponding $m_s$ simultaneously. Importantly, $D_{S(\pm 1/2)}$–$D_{T(\pm 1/2)}$ form an encoded spin qubit controllable by exchange interactions[9,23,32]. Due to the spin selection rule, the exchange interaction only couples the states with the same $S_{tot}$ and $m_s$ [9], thereby facilitating effective qubit operations.

The eigenspectrum of the Heisenberg trimer is altered by finite $\Delta B_z$, which hybridizes $D_{T(\pm 1/2)}$ and $Q_{(\pm 1/2)}$ states. In the $|\Delta B_z| \lesssim J_{LR}$ regime, such hybridization infers the

$\Delta B_z$-induced-perturbation of $D_{T(\pm 1/2)}$ energy, which modulates the $D_{S(\pm 1/2)}$–$D_{T(\pm 1/2)}$ energy gap $\Delta E(D_{T(\pm 1/2)})=E(D_{T(\pm 1/2)})-E(D_{S(\pm 1/2)})$. Here, $E(A_{(m_s)})$ denotes the energy of $A_{(m_s)}$. However, when $|\Delta B_z| \gg J_{LR}$, $\Delta B_z$ significantly transforms the eigenspectrum. For example, in the $m_s=+1/2$ subspace under $|\Delta B_z| \gg J_{LR}$, $D_{T0(+1/2)}=|T_0\rangle|\uparrow\rangle$ and $D_{T1(+1/2)}=|T_+\rangle|\downarrow\rangle$, whose energies are separated by $\Delta B_z$, become the eigenstates of Eqn. 1 (inset to Fig. 1b) instead of $D_{T(+1/2)}$ and $Q_{(+1/2)}$ for $|\Delta B_z| \lesssim J_{LR}$. Due to the hybridization, both $D_{T0(\pm 1/2)}$ and $D_{T1(\pm 1/2)}$ are exchange-coupled to $D_{S(\pm 1/2)}$, where the energy separation given by $\Delta B_z$ allows selective addressing of each state. As detailed below, $D_{T0(\pm 1/2)}$ and $D_{S(\pm 1/2)}$ form a "true" DFS decoupled from both short- and long-wavelength magnetic field noise.

The DFS states couple equally to certain degrees of environmental fluctuations; as a result, the energy gaps between these states remain invariant to the fluctuations[11,13]. For instance, in our system, long-wavelength magnetic field fluctuations $\delta B_{long}$ (Fig. 1c, blue curve), which uniformly affect the spins in the left and right sites, equally shift the energies of the states in the same $m_s$ subspace[9,13]. In other words, the trimer states with the same $m_s$ form a DFS protected from $\delta B_{long}$ whose characteristic length scale is longer than $d_{LR}$. In this regard, the $D_{S(\pm 1/2)}$–$D_{T(\pm 1/2)}$ ($D_{S(\pm 1/2)}$–$D_{T0(\pm 1/2)}$) states form a DFS, where the corresponding energy gap $\Delta E(D_{T(\pm 1/2)})$ ($\Delta E(D_{T0(\pm 1/2)})=E(D_{T0(\pm 1/2)})-E(D_{S(\pm 1/2)})$) is resilient to $\delta B_{long}$ as illustrated in the left panel in Fig. 1d (Fig. 1e).

In contrast, the short-wavelength magnetic field fluctuation $\delta B_{short}$ (Fig. 1c, red curve) influences the two sites of the trimer inhomogeneously and perturbs $\Delta B_z$ in addition to the average magnetic field across the sites. Because $\Delta E(D_{T(\pm 1/2)})$ is perturbed by $\Delta B_z$, $\Delta E(D_{T(\pm 1/2)})$ is prone to $\delta B_{short}$, and only $\Delta E(D_{T0(\pm 1/2)})$ remains constant also for $\delta B_{short}$. To elucidate this, we first introduce the quantity $\langle S_L^z \rangle_A = \langle S_1^z + S_2^z \rangle_A = \langle A|S_1^z + S_2^z|A\rangle$

($<S_R^z>_A=<S_3^z>_A=<A|S_3^z|A>$), which quantifies the average z-axis spin value of the spins residing in the left (right) site for a state $A$. For instance, $D_{S(+1/2)}=|S\rangle|\uparrow\rangle$, and $D_{T0(+1/2)}=|T_0\rangle|\uparrow\rangle$ have $<S_L^z>_{DS(+1/2)}=<S_L^z>_{DT0(+1/2)}=0$ and $<S_R^z>_{DS(+1/2)}=<S_R^z>_{DT0(+1/2)}=+1/2$; that is, $\delta<S_L^z>_{DT0(+1/2)}=<S_L^z>_{DT0(+1/2)}-<S_L^z>_{DS(+1/2)}=0$ and $\delta<S_R^z>_{DT0(+1/2)}=<S_R^z>_{DT0(+1/2)}-<S_R^z>_{DS(+1/2)}=0$. $\delta<S_L^z>_{A(ms)}=\delta<S_R^z>_{A(ms)}=0$ implies that $A_{(ms)}$ and $D_{S(ms)}$ shift equally in energy with respect to the local magnetic field fluctuation in the left and right sites, and therefore, the energy gap is decoupled from $\delta B_{short}$. In this regard, $\delta<S_L^z>_{A(ms)}$ ($\delta<S_R^z>_{A(ms)}$) quantifies the susceptibility of the $A_{(ms)}$-$D_{S(ms)}$ gap to the local magnetic fluctuation in the left (right) site. However, because $D_{T(+1/2)}$ ($D_{T(-1/2)}$) consists of a coherent mixture of $|T_0\rangle|\uparrow\rangle$ ($|T_0\rangle|\downarrow\rangle$) and $|T_+\rangle|\downarrow\rangle$ ($|T_-\rangle|\uparrow\rangle$), both $\delta<S_L^z>_{DT(\pm1/2)}$ and $\delta<S_R^z>_{DT(\pm1/2)}$ are non-zero, implying that $\Delta E(D_{T(\pm1/2)})$ is vulnerable to $\delta B_{short}$.

In Fig. 1f, we illustrate the eigenspectrum of the trimer as a function of $\Delta B_z/J_{LR}$ calculated from Eqn. 1 with $J_{LR}=J_{23}=J_{31}=0.1J_{12}$. As expected, $\Delta E(D_{T(\pm1/2)})$ exhibits a finite slope about $\Delta B_z$ i.e. $|\partial\Delta E(D_{T0(\pm1/2)})/\partial\Delta B_z|>0$ for $\Delta B_z<J_{LR}$. However, as $\Delta B_z$ is increased, $\Delta E(D_{T0(\pm1/2)})$ converges to $J_{12}$, and $\partial\Delta E(D_{T0(\pm1/2)})/\partial\Delta B_z$ approaches 0, which implies protection against $\delta B_{short}$. Analytically, $\Delta E(D_{T0(\pm1/2)}) \sim J_{12}\pm J_{LR}^2/(2\Delta B_z)+O(1/\Delta B_z^2)$ is derived from Eqn. 1 for $\Delta B_z \gg J_{LR}$, illustrating that the energy gap is insensitive to $\Delta B_z$ to the first order. We further show $\delta<S_L^z>$ ($\delta<S_R^z>$) for the different trimer eigenstates as a function of $\Delta B_z/J_{LR}$ in Fig. 1g (Fig. 1h). As expected, $\delta<S_L^z>_{DT0(\pm1/2)}$ and $\delta<S_R^z>_{DT0(\pm1/2)}$ approach 0 for large $\Delta B_z/J_{LR}$ (red shaded box), which reflects decoupling from $\delta B_{short}$.

**Heisenberg spin-1/2 trimer in a semiconductor double quantum dot**

Experimentally, we utilize a gate-defined semiconductor DQD to host the aforementioned Heisenberg spin-1/2 trimer. Figure 2a shows a scanning electron micrograph of the GaAs QD device employed in this work[23] (see Methods). The device was placed in a dilution refrigerator with a base temperature of 20 mK with a resulting electron temperature of $T_e$~230 mK. A variable in-plane magnetic field $B_{ext}$ was applied in the direction indicated by the white arrow. With the two-dimensional electron gas residing ~70 nm below the surface, QDs can be defined by the gate voltages[33]. In this work, we focus on the right DQD (green dots) and the single-electron transistor (SET) charge sensor on the right (yellow dot) by grounding the irrelevant gates. With this SET, we perform the high-bandwidth radio frequency (rf) reflectometry detection of the DQD charge state[34].

When two particles are confined in a QD, the Coulomb correlation between the particles become more significant in a larger QD[19,21,22]. While the orbital splitting of a QD is usually given by the confinement energy, a strong Coulomb correlation in a large QD renormalizes the orbital states and quenches the orbital splitting (or singlet-triplet splitting) of the QD[19,21,22]. Additionally, anisotropic QD confinement promotes Wigner molecularizations and further decreases the orbital splitting[22]. In our DQD, to time-resolve the electron tunneling events between the right QD and its reservoir, the QD is tuned less transparent compared to the left one implying that the left QD has a larger QD radius. Moreover, we find the left QD confinement is anisotropic[23], which fosters Wigner-molecularization and leads to small orbital splitting $\delta L_{ST} < 10$ GHz·$h$. In contrast, a negligible correlation effect in the right QD results in a large orbital splitting of the QD, $\delta R_{ST}$ ~108 GHz·$h$ (right panel in Fig. 2a), which is estimated by pulsed-gate spectroscopy[35] (see Supplementary Note S1).

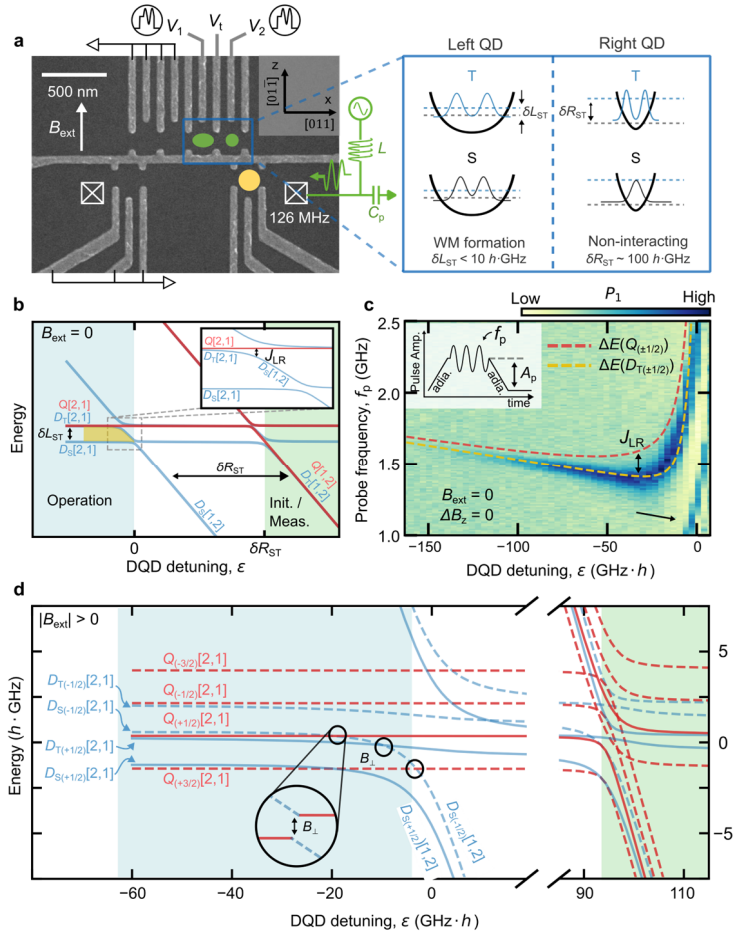

**Fig. 2 a,** Scanning electron microscopy of the GaAs quantum dot (QD) device used in this work. The green dots denote the double QD (DQD) hosting a three-electron Heisenberg trimer aligned along the [011] crystal axis. As illustrated in the right panel, the left (right) QD confinement potential promotes (prevents) the formation of a Wigner molecule (WM), resulting in a small (large) orbital splitting $\delta L_{ST}$ ($\delta R_{ST}$) in the left (right) QD. The yellow dot depicts a radio-frequency single-electron transistor (rf-SET) charge sensor. An LC tank circuit is attached to the ohmic contact close to the rf-SET. An external magnetic field $B_{ext}$ is applied along the white arrow. **b,** Eigenenergy spectrum of a three-electron system in a DQD at $B_{ext} = \Delta B_z = 0$. The corresponding energies along the [2,1]-[1,2] charge configurations are presented as a function of DQD detuning, $\varepsilon$, with small (large) $\delta L_{ST}$ ($\delta R_{ST}$). Inset: Zoom-in of the energy diagram near $\varepsilon = 0$. $J_{LR}$ represents the inter-QD exchange interaction. **c,** Pulsed-microwave spectroscopy of the trimer at [2,1] at $B_{ext} = 0$. The probability of the excited orbital state, $P_1$ is shown as a function of the applied microwave probe frequency $f_p$, and $\varepsilon$ (see inset for the pulse schematic. $A_p$ is the amplitude of the detuning pulse.). The yellow (red) dashed curves illustrate the $D_{T(\pm 1/2)}$-$D_{S(\pm 1/2)}$ ($Q_{(\pm 1/2)}$-$D_{S(\pm 1/2)}$) energy gap calculated from the three-electron Hamiltonian (Supplementary Note S2). The black arrow denotes the transition driven by the two-photon process. **d,** Eigenspectrum at $|B_{ext}| > 0$ as a function of $\varepsilon$. The black circles denote avoided crossings between states with different $m_s$ due to the finite transverse nuclear Overhauser field $B^\perp$. Inset: Zoom-in of the $B^\perp$ avoided-crossing between $D_{S(-1/2)}[2,1]$ and $Q_{(+1/2)}[2,1]$.

The DQD is operated in the [2,1]–[1,2] charge configuration to define the Heisenberg spin-1/2 trimer. Here, n (m) denotes the number of electrons in the left (right) QD by [n,m] notation. Figure 2b presents the relevant energy diagram as a function of the DQD detuning $\varepsilon$, at $B_{ext}=\Delta B_z=0$ (We naturally adopt $\Delta B_z=B_z^L–B_z^R$ from Fig. 1, defining $B_z^L$ ($B_z^R$) as the local magnetic field at the left (right) QD.). In a deep-detuned [2,1] or [1,2] regime, where the direct charge transitions between the QDs are inhibited, the system can be described by the anti-ferromagnetic Heisenberg Hamiltonian presented in Eqn. 1 [9]. In particular, the [2,1] configuration is identical to the case presented in Fig. 1a with $J_{12} \gg J_{LR}$ where $J_{12}$ is now dominated by $\delta L_{ST}$. Thereby in [2,1], when two electrons in the left QD form a spin-singlet and occupy the ground orbital, the trimer state becomes a $D_{S(\pm 1/2)}[2,1]$. $A_{(ms)}[n,m]$ denotes the corresponding spin and DQD charge state simultaneously. However, if the two electrons in the left QD form a spin-triplet and occupy the excited orbital, the trimer state can be either a $D_{T(\pm 1/2)}[2,1]$ or a quadruplet state ($Q_{(\pm 1/2)}[2,1]$ or $Q_{(\pm 3/2)}[2,1]$) depending on $S_{tot}$.

At $\varepsilon > \delta R_{ST}$ (green shaded region in Fig. 2b) if the Fermi-level of the reservoir lies in between the ground and excited orbital levels, only the electron from an excited orbital ($D_{T(\pm 1/2)}[1,2]$, $Q_{(\pm 1/2)}[1,2]$, or $Q_{(\pm 3/2)}[1,2]$) can tunnel out to the reservoir and tunnel back in to fill the ground state $D_{S(\pm 1/2)}[1,2]$, known as the energy-selective tunneling (EST) process[23,36,37]. With $\delta R_{ST} \sim 108$ GHz·$h > 50 k_B T_e$ ($k_B$ is the Boltzmann constant), the rf-SET in our device allows the time-resolved detection of such EST, allowing the single-shot measurement of the trimer states[23].

Based on the EST readout, we explore the excitation spectrum of the trimer in the [2,1] configuration at $B_{ext}=\Delta B_z=0$ by pulsed-microwave spectroscopy[38]. For the spectroscopy, we first initialize the spin state to $D_{S(\pm 1/2)}[1,2]$ at $\varepsilon_{init} > \delta R_{ST}$ exploiting the EST[23]. By pulsing the initialized state into the [2,1] configuration adiabatically with respect to the inter-dot tunnel coupling rate, we transfer $D_{S(\pm 1/2)}[1,2]$ to $D_{S(\pm 1/2)}[2,1]$. Then, we apply an electric microwave tone to gate $V_1$ (Fig. 2a), which excites $D_{S(\pm 1/2)}[2,1]$ to $D_{T(\pm 1/2)}[2,1]$ if the frequency of the tone matches $\Delta E(D_{T(\pm 1/2)})/h$. The $D_{T(\pm 1/2)}[2,1]$ population can be detected via the EST process, once adiabatically brought back to $D_{T(\pm 1/2)}[1,2]$ at $\varepsilon_{init}$. Figure 2c shows the resulting $D_{T(\pm 1/2)}[2,1]$ population as a function of the microwave frequency $f_p$ and $\varepsilon$, which reveals the degenerate spectrum corresponding to $\Delta E(D_{T(+1/2)}[2,1])=\Delta E(D_{T(-1/2)}[2,1])$. The pulse amplitude $A_p=\varepsilon_p-\varepsilon_{init}$ (schematic in Fig. 2c) is varied to sweep over $\varepsilon$. Here, the pulse-tip $\varepsilon_p$ is the $\varepsilon$ reached by the pulse. The yellow (red) dashed curve superposed onto Fig. 2c corresponds to $\Delta E(D_{T(\pm 1/2)}[2,1])$ ($\Delta E(Q_{(\pm 1/2)}[2,1])=\Delta E(Q_{(\pm 3/2)}[2,1])$) calculated from our three-electron DQD Hamiltonian (see Supplementary Note S2). We empirically find $\delta L_{ST}\sim 1.3$ GHz·h, which is an order of magnitude smaller than the orbital splitting $\sim 70$ GHz·h expected from the lithographic size of the QD. This confirms the Wigner-molecularization in the left QD. The transitions from $D_{S(\pm 1/2)}[2,1]$ to $Q_{(\pm 1/2)}[2,1]$ or to $Q_{(\pm 3/2)}[2,1]$ (red dashed curve in Fig. 2c) are prohibited due to the spin selection rule[9]; hence, these transitions are not visible on the measured spectrum. Furthermore, the energy difference between $Q_{(\pm 1/2)}[2,1]$ and $D_{T(\pm 1/2)}[2,1]$ corresponds to the size of $J_{LR}$ which varies with $\varepsilon$. Below, only the trimer states in the [2,1] configuration are discussed and the charge states are omitted from the notation.

For $|B_{ext}|>0$, the $D_S$ and $D_T$ states split into $m_s=-1/2, +1/2$, and the $Q$ state splits into $m_s = -3/2, -1/2, +1/2, +3/2$ states (Fig. 2d). When the states with different $m_s$ become degenerate due to Zeeman splitting, the degeneracy is lifted by the transverse nuclear Overhauser field $B^\perp$ and results in an avoided-crossing between the states with $|\Delta m_s| = 1$ [24,39]. These avoided-crossings are marked by black circles in Fig. 2d. Finite $B^\perp$ allows DNP processes which also builds $\Delta B_z>0$ [24,39,40] (see Supplementary Note S3).

## Spectroscopy of the spin trimer under finite $\Delta B_z$

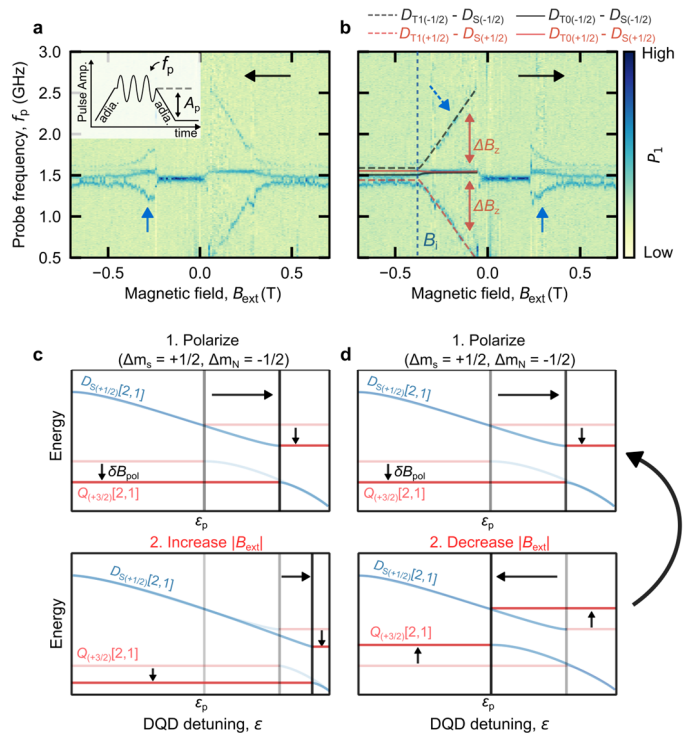

**Fig. 3 a (b),** Magnetospectroscopy of the trimer with decreasing (increasing) $B_{ext}$. Excited orbital state probability $P_1$ is recorded as a function of $f_P$ and $B_{ext}$, with constant $A_p$ (see inset to **a** for the pulse schematics). The black arrow illustrates the $B_{ext}$ sweep direction. Pulse-induced DNP (see Supplementary Note S3) builds hysteretic $\Delta B_z$ about the sweep direction and leads to the four different transitions presented. The black (red) solid and dashed lines represent the $D_{T0}$–$D_S$ and $D_{T1}$–$D_S$ energy splitting in $m_s = -1/2$ ($m_s = 1/2$) subspace, respectively, calculated from the toy-model Hamiltonian together with varying $\Delta B_z$ (see Supplementary Note S2). **c,** Schematic of the pulse-induced DNP process for increasing $|B_{ext}|$. Top panel: when the pulse tip, $\varepsilon_p$, coincides with the $B^\perp$ avoided-crossing, DNP is facilitated to increase the electron Zeeman splitting by $\delta B_{pol}$, and shifts the avoided-crossing to a more positive $\varepsilon$. Bottom panel: Increasing $|B_{ext}|$ shifts the avoided-crossing to more positive $\varepsilon$, prohibiting additional DNP. This results in the DNP in a limited region (solid blue arrows in **a** and **b**). **d,** Schematic of the pulsed induced DNP process for decreasing $|B_{ext}|$. Top panel: the DNP shifts the $B^\perp$ avoided-crossing to a more positive $\varepsilon$ as shown in **c**. Bottom panel: Decreasing $|B_{ext}|$ refocuses the avoided-crossing to $\varepsilon_p$, which enables additional DNP.

Figures 3a,b show the magnetospectroscopy of the trimer with different $B_{ext}$ sweep directions. For the spectroscopy, a microwave tone is superposed to the polarization phase of the DNP pulse shown in Supplementary Note S3, implying the pulse is also capable of DNP.

Here, we keep the pulse detuning amplitude $A_p \sim 200$ GHz·h to reach $\varepsilon = \varepsilon_p \sim -70$ GHz·h in the [2,1] configuration. Here, $\Delta E[D_{T(\pm 1/2)}] \sim J_{12} \sim 1.45$ GHz·h and $J_{LR} \sim 0.15$ GHz·h present the relevant energy scales. As $B_{ext}$ is increased from 0 to 231.5 mT to match $(g^*\mu_B)B_{ext} = \Delta E[D_{T(\pm 1/2)}]$ (Fig. 3b), from which we extract $|g^*| \sim 0.44$ [41], $D_{S(-1/2)}$ becomes degenerate with $D_{T(+1/2)}$ at $\varepsilon_p$ due to the Zeeman splitting, and DNP occurs. As a consequence, the finite $\Delta B_z$ due to the DNP mixes $D_{T(+1/2)}$ ($D_{T(-1/2)}$) and $Q_{(+1/2)}$ ($Q_{(-1/2)}$) where $D_{S(+1/2)}$ ($D_{S(-1/2)}$) can now be electrically excited to both mixtures as discussed above. This explains the four transitions viable in the spectrum for $B_{ext} > \Delta E[D_{T(\pm 1/2)}]$.

The DNP yields a highly hysteretic spectrum depending on the field sweep direction. If $\varepsilon_p$ coincides with a $B^{\perp}$ avoided-crossing, the DNP increases the size of the Zeeman splitting which shifts the avoided-crossing to more positive $\varepsilon$ (Fig. 3c, top panel). A further increase in $|B_{ext}|$ in the next sweep step shifts the crossing to an even more positive $\varepsilon$, prohibiting additional DNP (Fig. 3c, bottom panel). Consequently, while increasing $|B_{ext}|$, the pulse can induce DNP within a small $B_{ext}$ window (near the blue solid arrows in Fig. 3a,b). This results in a small $\Delta B_z$, which eventually decays as a function of time. In contrast, when decreasing $|B_{ext}|$, even though the DNP shifts the avoided-crossing towards positive $\varepsilon$ (Fig. 3d, top panel), decreasing $|B_{ext}|$ re-focuses the crossing to $\varepsilon_p$, allowing additional DNP (Fig. 3d, bottom panel). We note that such re-focusing is possible because the relaxation time of the Overhauser field ($\sim 10^0$ mins [24]) is longer than the time required to decrease $|B_{ext}|$ by one sweep step ($\sim 10$ s). In other words, when decreasing $|B_{ext}|$, the DNP translates the decreased $B_{ext}$ into the Overhauser field to preserve the constant Zeeman splitting experienced by the electrons, and maintains the $B^{\perp}$ avoided-crossing at $\varepsilon_p$. This allows the DNP to persist over a large range of $B_{ext}$, and yields large $\Delta B_z \sim 1.05$ GHz·h (red arrows in Fig. 3b), corresponding to $\Delta B_z/J_{LR} \sim 6.67$. Moreover, although the quenching of $\Delta B_z$ at low $|B_{ext}|$ suggests that a minimum $|B_{ext}|$ is required to maintain the polarization of the nuclear spins, detailed investigations are required to verify the quenching mechanism.

The red (black) solid and dashed lines in Fig. 3b denote the energy splittings in the $m_s=+1/2$ ($m_s=-1/2$) subspace calculated from the toy-model Hamiltonian with finite $\Delta B_z$. For the calculation, $\Delta B_z \sim 50$ MHz is assumed in the range $B_{ext}= -700$ mT $\sim B_i$, where we suppose that DNP becomes effective for $|B_{ext}|<|B_i|$ to linearly build $\Delta B_z$ up to $\sim 1.05$ GHz·$h$. Noting that $B_{ext}$ is translated into the Overhauser field when decreasing $|B_{ext}|$, the slope $\sim 0.53$ (denoted by the blue-dashed arrow in Fig. 3b) infers the effectiveness of our DNP in building $\Delta B_z$. As illustrated in Fig. 1f, the excitation spectra in the $m_s = \pm 1/2$ subspace asymptotically approach $\Delta E[D_{T0(\pm 1/2)}]$ and $\Delta E[D_{T1(\pm 1/2)}]$ for large $\Delta B_z$, reaching the "true" DFS formed by $D_{S(\pm 1/2)}$ and $D_{T0(\pm 1/2)}$.

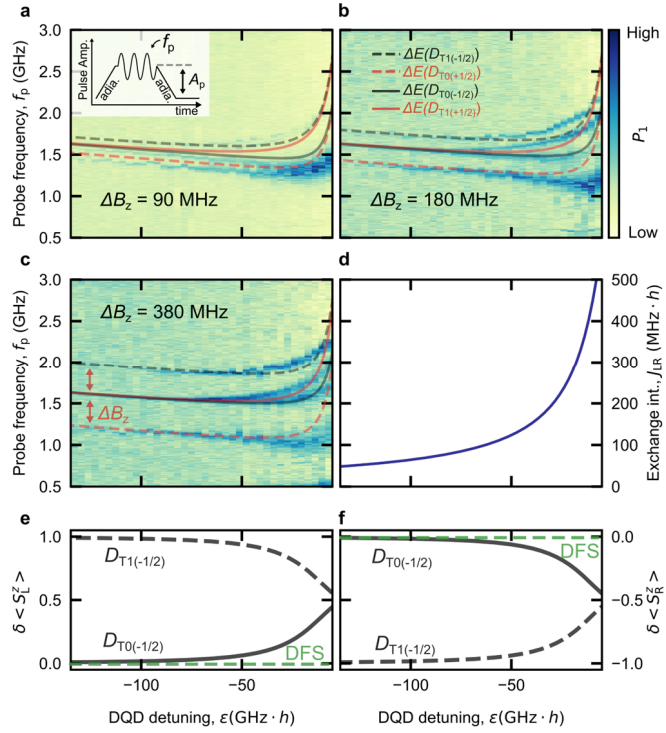

**Fig. 4 a-c,** Pulsed microwave spectroscopy of the trimer along the DQD detuning $\varepsilon$ for $\Delta B_z$ = 90, 180, and 380 MHz·$h$. Excited orbital state population $P_1$ is recorded as a function of the microwave probe frequency $f_p$ and $\varepsilon$ for each panel. The superposed lines correspond to the energy spectrum calculated from the toy-model Hamiltonian with the respective $\Delta B_z$ (see Supplementary Note S2). Inset to **a**: Schematic of the pulse utilized for the spectroscopy. **d,** Inter-QD exchange interaction strength $J_{LR}$ as a function of $\varepsilon$, evaluated from the Hamiltonian. **e (f),** Susceptibility of the energy gap to magnetic field fluctuation in the left (right) QD as a function of $\varepsilon$. For the calculation, $J_{LR}(\varepsilon)$ shown in **d** is assumed with $\Delta B_z$ = 380 MHz·$h$. $\delta\langle S_L^z\rangle = 0$ ($\delta\langle S_R^z\rangle = 0$), represented by the green dashed line, implies the decoupling of the relevant energy gaps from the local magnetic field fluctuation in the left (right) QD.

Although Fig. 3 readily demonstrates the modulation of the eigenspectrum under $\Delta B_z \gg J_{LR}$, we additionally vary $J_{LR}$ with $\varepsilon$ and show the resulting eigenspectrum in Fig. 4. Figure 4a–c present the pulsed microwave spectra of the trimer spanned by $\varepsilon$ and $f_P$, at different $\Delta B_z$. To generate finite $\Delta B_z$, we perform the magnetospectroscopy shown in Fig. 3a,b by decreasing $|B_{ext}|$ from a large value and stopping at a certain value. From our toy-model Hamiltonian, we calculate the eigenspectrum as a function of $\varepsilon$ at constant $\Delta B_z$, plotted in Fig.4a–c (see Supplementary Note S2), where we estimate $\Delta B_z$ = 80 (180 and 380) MHz·$h$ for Fig. 4a (4b,c) by empirically fitting to the eigenspectrum. We note the simulated spectra may deviate from the observations due to the nuclear spin diffusion during the measurements[42].

As shown in Fig. 4d, $J_{LR}(\varepsilon)$ decays as $\varepsilon$ is more negatively detuned (Fig. 2b,c). Consequently, for a fixed $\Delta B_z$, the ratio $\Delta B_z/J_{LR}$ increases for negative detuning, where the "true" DFS emerges. We further evaluate $\delta\langle S_L^z\rangle$ ($\delta\langle S_R^z\rangle$) in Fig. 4e (4f) as a function of $\varepsilon$, assuming $\Delta B_z$ = 380 MHz·$h$. Evidently, both $\delta\langle S_L^z\rangle_{DT0(-1/2)}$ and $\delta\langle S_R^z\rangle_{DT0(-1/2)}$ approach 0 for large detuning, implying protection against $\delta B_{short}$.

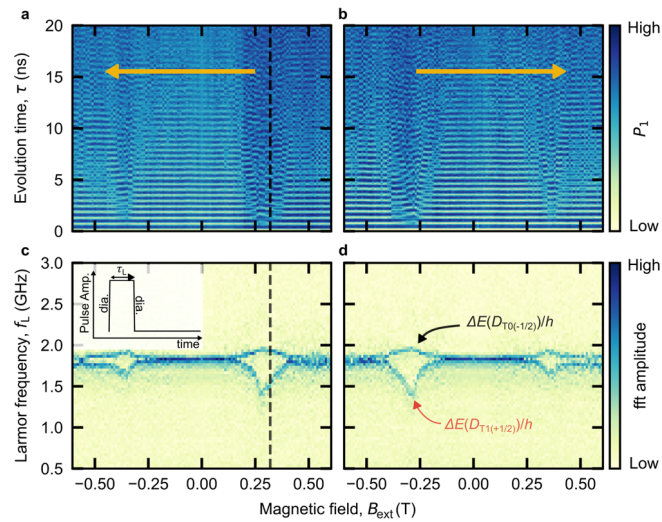

**Fig. 5 a (b),** Landau–Zener–Stückelberg (LZS) oscillation of the trimer with decreasing (increasing) $B_{ext}$. Excited orbital state population $P_1$ is recorded as a function of the free evolution time $\tau$ (see inset to **c** for the pulse schematic), and $B_{ext}$. **c (d),** Fast-Fourier-transformation (FFT) of the LZS oscillation shown in **a (b)**. The LZS pulse-induced DNP builds finite $\Delta B_z$ and results in the two distinct frequencies related to $\Delta E(D_{T0(-1/2)})$ and $\Delta E (D_{T1(+1/2)})$. The Larmor frequencies corresponding to $\Delta E(D_{T0(+1/2)})$ and $\Delta E (D_{T1(-1/2)})$ are not visible due to the small Landau-Zener transition given by the finite rise time of the pulse.

We further probe the trimer by coherent Landau-Zener-Stückelberg (LZS) oscillation with different $B_{ext}$ sweep directions (Fig. 5a,b). The corresponding fast-Fourier-transformations (FFTs) are shown in Fig. 5c,d, respectively. Here non-adiabatic square pulses with varying length $\tau$ and with a fixed amplitude are applied to induce LZS oscillation[24,43].

These LZS pulses can similarly result in DNP when the pulse tip matches the degeneracy as in Fig. 3a,b, which in turn reveals $\Delta B_z$ split states evident from the FFT.

Compared to the four clear transitions shown in Fig. 3a,b, however, two branches are faintly visible on the LZS spectrum even though two energy levels are accessible for each $D_{S(+1/2)}$ and $D_{S(-1/2)}$. We ascribe this to the negligible Landau-Zener transition probability to the upper excited state for each $m_s$ subspace due to the finite rise-time of the LZS pulses[44]. Explicitly, if the state is initialized to $D_{S(+1/2)}$ ($D_{S(-1/2)}$), the LZS oscillation mainly reveals the phase oscillation corresponding to $\Delta E(D_{T1(+1/2)})$ ($\Delta E(D_{T0(-1/2)})$), which is lower in energy than $\Delta E(D_{T0(+1/2)})$ ($\Delta E(D_{T1(-1/2)})$), as shown in Fig. 1f. Moreover, the effect of DNP is less prominent with this pulsing scheme compared to the case in Fig. 3a,b because average time pulse spends at the $B^{\perp}$ avoided-crossing is shorter in this measurement sequence.

# Real-time Bayesian estimation of the spin trimer energy splitting

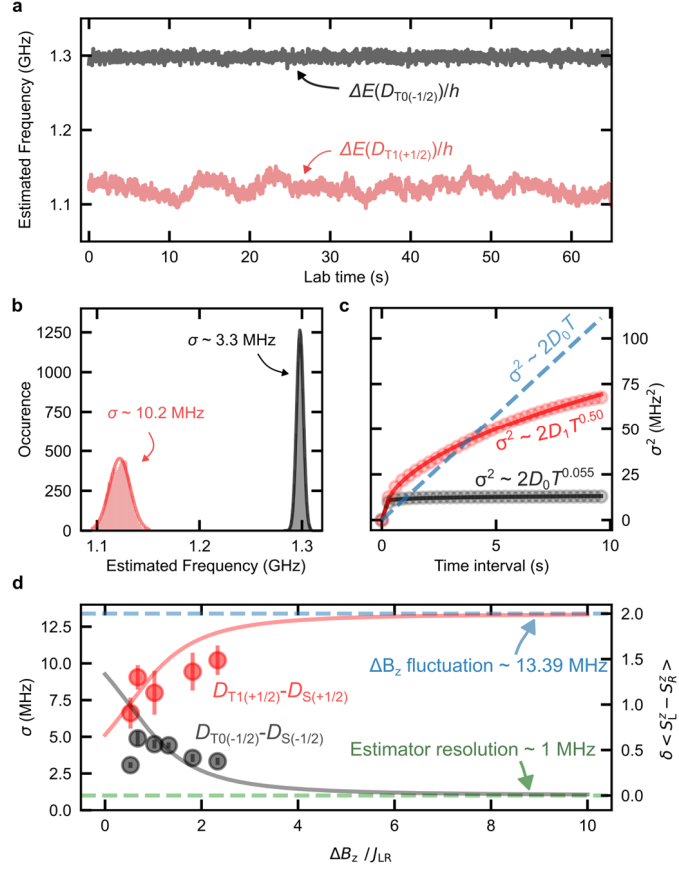

**Fig. 6 a,** Bayesian-estimated $\Delta E(D_{T0(-1/2)})/h$ [$\Delta E(D_{T1(+1/2)})/h$] along laboratory time shown in the black [red] trace. Both energy splittings are simultaneously estimated every 6 ms based on 300 EST single-shot measurements. **b,** Histogram of the frequency fluctuation shown in **a**. Numerical fit (solid line) to a Gaussian function yields the standard deviation σ ~ 3.3 MHz [10.2 MHz] for $\Delta E(D_{T0(-1/2)})/h$ [$\Delta E(D_{T1(+1/2)})/h$] fluctuation. **c,** Variance of the fluctuation σ² as a function of the time interval $T$. Numerical fit (solid line) to a sub-diffusion model σ² = 2D$T^α$ result in D = 5.78 MHz²/s$^{0.055}$ [11.19 MHz²/s$^{0.5}$] and α = 0.055 [0.5] for $\Delta E(D_{T0(-1/2)})/h$ [$\Delta E(D_{T1(+1/2)})/h$] fluctuation. The blue dashed line illustrates the expected variance for a standard diffusion process with D = 5.78 MHz²/s. **d,** Standard deviation of $\Delta E(D_{T0(-1/2)})/h$ [$\Delta E(D_{T1(+1/2)})/h$] measured for different $\Delta B_z/J_{LR}$ presented as solid black (red) circles. The lower (upper) bound of the fluctuation is set by the estimator resolution ~ 1 MHz (standard deviation of the $\Delta B_z$ fluctuation ~ 13.39 MHz, see Supplementary Note S4) shown as green (blue) dashed lines. The black [red] solid curve correspond to δ<$S_L^z$–$S_R^z$> of $D_{T0(-1/2)}$ [$D_{T1(1/2)}$] branch as a function of $\Delta B_z/J_{LR}$.

Based on the coherent LZS oscillation we further investigate the real-time fluctuation of the energy levels by Bayesian Hamiltonian estimation[29–31]. The estimator evaluates the two different frequencies corresponding to $\Delta E(D_{T0(-1/2)})/h$ and $\Delta E(D_{T1(+1/2)})/h$ simultaneously, based on 300 EST single-shot measurement outcomes (see Methods). Figure 6a shows the estimated frequencies as a function of the laboratory time. Here, each single-shot measurement takes ~ 20 μs, implying ~ 6 ms per estimation.

Figure 6b presents the histograms of the time traces shown in Fig. 6a. From the histogram we extract the standard deviations of $\Delta E(D_{T0(-1/2)})/h$ and $\Delta E(D_{T1(+1/2)})/h$, $\sigma_{DT0(-1/2)}$ and $\sigma_{DT1(+1/2)}$, respectively. Notably, $\sigma_{DT0(-1/2)}$ is significantly smaller than $\sigma_{DT1(+1/2)}$. This clearly demonstrates the protection against $\delta B_{short}$ for $\Delta E(D_{T0(-1/2)})$ in comparison to $\Delta E(D_{T1(+1/2)})$. Figure 6c additionally shows the variance, $\sigma^2$, along time interval $T$ over which the variance is evaluated[29,45]. Compared to the standard diffusion case (blue dashed line), where the variance increases linearly with $T$, the variances of $\sigma^2_{DT1(+1/2)}$ (red circles) and $\sigma^2_{DT0(-1/2)}$ (black circles) show $\sigma^2 \sim DT^\alpha$ with $\alpha<1$, and the diffusion coefficient D. This indicates the sub-diffusive behavior of the energy splittings, where we extract $\alpha=0.50$ ($\alpha=0.055$) for $\Delta E(D_{T1(+1/2)})/h$ ($\Delta E(D_{T0(-1/2)})/h$ ). The small $\alpha$ suggests that the states are mainly affected by a non-Gaussian noise correlator[46], implying the states are less affected by the nuclear spin diffusion, which is assumed to be a typical Gaussian noise correlator[29,42].

Time traces of the energy gaps at different $\Delta B_z$ and at different $J_{LR}(\varepsilon)$ are recorded to further probe the effect of $\Delta B_z/J_{LR}$ on the robustness. In Fig. 6d, which shows $\sigma_{DT0(-1/2)}$ and $\sigma_{DT1(+1/2)}$ at different $\Delta B_z/J_{LR}$, we superpose $\delta\langle S_L^z - S_R^z\rangle_{DT1(-1/2)} = \delta\langle S_1^z + S_2^z - S_3^z\rangle_{DT1(-1/2)}$ ($\delta\langle S_L^z - S_R^z\rangle_{DT0(1/2)}$) as a function of $\Delta B_z/J_{LR}$ as a black (red) curve. $\delta\langle S_L^z - S_R^z\rangle_{A(ms)}$ quantifies the degree of protection against $\Delta B_z$ noise for the $A_{(ms)}$–$D_{S(ms)}$ gap. Here, we assume that, for minimal $\delta\langle S_L^z - S_R^z\rangle = 0$, the frequency resolution of our estimator sets the lower bound of $\sigma_{min}$ ~1 MHz. In addition, $\Delta B_z$ fluctuation ~13.39 MHz, measured by the singlet-triplet qubit in this device (see Supplementary Note S4), sets the upper bound at the maximal $\delta\langle S_L^z -$

$S_R^z$>=2 [40]. Because the Bayesian estimation is carried out when $\Delta B_z$ <200 MHz·h, we expect smaller $\sigma_{DT0(-1/2)}$ for larger $\Delta B_z$, which is indeed within reach as demonstrated in Fig. 3a,b. Thereby, we anticipate a larger $\Delta B_z$ to provide enhanced protection against $\delta B_{short}$ in addition to the inherent $\delta B_{long}$ resilience, which would further facilitate highly coherent spin qubit operations completely decoupled from the magnetic field noise.

**Discussion**

In this work, we present a "true" DFS, robust to both short- and long-wavelength magnetic field fluctuations, in an antiferromagnetic Heisenberg spin-1/2 trimer. Compared to previously proposed "true" DFSs schemes in QDs[16,17], the DFS presented here offers a more compact and practical route toward magnetic noise-free spin qubit operations. We envision a well-engineered pulse scheme, altering between the DNP and control phase, to facilitate coherent qubit control in the DFS in GaAs. Also, the integration of a micromagnet[47] or the site-dependent g-factors of QDs in planar Ge may provide an efficient means to generate a sizable Zeeman splitting difference[48]. This alleviates the need for DNP phases to maintain $\Delta B_z$ and allows simpler qubit control schemes in the presented DFS, which may promote the operations of multiple qubits residing in the DFS. Additionally, exploiting the tunability of the strongly-correlated states[24,27] the dispersion of the energy levels can be tuned to have a smaller slope about $\varepsilon$, which may further suppress the charge noise susceptibility. Furthermore, the DFS studied in this work could be realized in other exchange-coupled spin systems in general[49,50], suggesting an alternative route toward quantum error-mitigation in various platforms.

**Methods**

Device fabrication

A quadruple QD device was fabricated on a GaAs/AlGaAs heterostructure with a two-dimensional electron gas (2DEG) formed ~70 nm below the surface. An electronic mesa near the QD array was defined by wet etching leaving the 2DEG only around the QDs. A metal stack of Ni/Ge/Au was thermally diffused into the mesa to form the ohmic contacts. Metal gates for QD formation were defined by the standard electron-beam lithography and evaporation technique (5 nm Ti / 30 nm Au). Course gate structures including the bonding pads were deposited using photolithography and evaporation (5 nm Ti / 200 nm Au).

Eigenspectrum of the Heisenberg spin trimer at different $\Delta B_z$

We show the explicit spin expressions for the eigenstates of the Heisenberg spin trimer (Eqn. 1) at $\Delta B_z = 0$ (Table 1) and at $|\Delta B_z| \gg J_{LR}$ (Table 2). In each table, the state in the first (second) bracket refers to the two- (single-) spin state in the left (right) site. The spin triplet states $T_0$, $T_+$, $T_-$ correspond to $(|\uparrow\downarrow\rangle + |\downarrow\uparrow\rangle)/\sqrt{2}$, $|\uparrow\uparrow\rangle$, and $|\downarrow\downarrow\rangle$, respectively, and S denotes the spin singlet state, $(|\uparrow\downarrow\rangle - |\downarrow\uparrow\rangle)/\sqrt{2}$.

**Table 1 Eigenstates of the Heisenberg spin trimer at $\Delta B_z = 0$**

| State | Spin expression |
|---|---|
| $Q_{(+3/2)}$ | $|T_+\rangle|\uparrow\rangle$ |
| $Q_{(+1/2)}$ | $\frac{1}{\sqrt{3}}[\sqrt{2}|T_0\rangle|\uparrow\rangle + |T_+\rangle|\downarrow\rangle]$ |
| $Q_{(-1/2)}$ | $\frac{1}{\sqrt{3}}[\sqrt{2}|T_0\rangle|\downarrow\rangle + |T_-\rangle|\uparrow\rangle]$ |
| $Q_{(-3/2)}$ | $|T_-\rangle|\downarrow\rangle$ |

| | |
|---|---|
| $D_{S(+1/2)}$ | $\|S\rangle\|\uparrow\rangle$ |
| $D_{T(+1/2)}$ | $\frac{1}{\sqrt{3}}\left[\|T_0\rangle\|\uparrow\rangle - \sqrt{2}\|T_+\rangle\|\downarrow\rangle\right]$ |
| $D_{S(-1/2)}$ | $\|S\rangle\|\downarrow\rangle$ |
| $D_{T(-1/2)}$ | $\frac{1}{\sqrt{3}}\left[\|T_0\rangle\|\downarrow\rangle - \sqrt{2}\|T_-\rangle\|\uparrow\rangle\right]$ |

**Table 2** Eigenstates of the Heisenberg spin trimer at $|\Delta B_z| \gg J_{LR}$

| State | Spin expression |
|---|---|
| $Q_{(+3/2)}$ | $\|T_+\rangle\|\uparrow\rangle$ |
| $Q_{(-3/2)}$ | $\|T_-\rangle\|\downarrow\rangle$ |
| $D_{S(+1/2)}$ | $\|S\rangle\|\uparrow\rangle$ |
| $D_{T0(+1/2)}$ | $\|T_0\rangle\|\uparrow\rangle$ |
| $D_{T1(+1/2)}$ | $\|T_+\rangle\|\downarrow\rangle$ |
| $D_{S(-1/2)}$ | $\|S\rangle\|\downarrow\rangle$ |
| $D_{T0(-1/2)}$ | $\|T_-\rangle\|\downarrow\rangle$ |
| $D_{T1(-1/2)}$ | $\|T_-\rangle\|\uparrow\rangle$ |

Bayesian estimation

To record the fluctuation of the Heisenberg spin trimer eigenstates, we perform Bayesian Hamiltonian estimation[29–31] based on energy selective tunneling (EST) single-shot state detection[23,36,37]. For the estimation, we apply N = 300 different Landau-Zener-Stuckelberg (LZS) pulses (see Fig. 5) with the n$^{th}$ pulse having the free evolution time $\tau_n = n \cdot t_1 = n \cdot 0.12$ ns, where each pulse is followed by a single-shot readout. Because the LZS pulse mainly reveals the coherent oscillation between $D_{S(+1/2)}$–$D_{T1(+1/2)}$ and $D_{S(-1/2)}$–$D_{T0(-1/2)}$ (Fig. 5) the probability of measuring the excited (ground) state at the n$^{th}$ measurement is given by $P(M_n = e \mid f_1, f_2) = \alpha + \beta_1 \sin(2\pi f_1 \tau_n) + \beta_2 \sin(2\pi f_2 \tau_n)$ ($P(M_n = g \mid f_1, f_2) = 1 - P(M_n = e \mid f_1, f_2)$). Here $f_1$ and $f_2$ are the oscillation frequencies corresponding to the $D_{S(+1/2)}$–$D_{T1(+1/2)}$ and $D_{S(-1/2)}$–$D_{T0(-1/2)}$ oscillations, respectively, and $M_n$ is the n$^{th}$ measurement result which can be either the excited (e) or ground (g) state. P(A|B) refers to the conditional probability of A given B. According to the Bayes theorem, $P(f_1, f_2 \mid M_n) = P(M_n \mid f_1, f_2) P(f_1, f_2)/P(M_n)$, where a uniform distribution of $P(f_1, f_2)$ over the $f_1$ and $f_2$ space is supposed for simplicity. Assuming each measurement is not affected by the previous ones, $P(f_1, f_2 \mid M_1, M_2, \cdots, M_N) = \prod_{n=1}^{N} P(f_1, f_2 \mid M_n) \propto \prod_{n=1}^{N} P(M_n \mid f_1, f_2)$ holds.

For the efficient real-time estimation of the frequencies, we utilize a field-programmable-gate-array (FPGA, Digilent Zedboard) to update the probability distribution as soon as each measurement result is sent to the FPGA. After 300 measurements, the FPGA outputs a single set of ($f_1^*$, $f_2^*$) which maximizes $P(f_1, f_2 \mid M_1, M_2, \cdots, M_N)$. Thereby, a single estimation of the set takes ~ 6 ms and allows both the $D_{S(+1/2)}$–$D_{T1(+1/2)}$ and $D_{S(-1/2)}$–$D_{T0(-1/2)}$ fluctuations to be tracked simultaneously.

**Acknowledgments**

This work was supported by a National Research Foundation of Korea (NRF) grant funded by the Korean Government (MSIT) (No. 2019M3E4A1080144, No. 2019M3E4A1080145, No. 2019R1A5A1027055, No. RS-2023-00283291, SRC Center for Quantum Coherence in Condensed Matter No. RS-2023-00207732, No. RS-2024-00442994, No. RS-2024-00413957, and No. 2023R1A2C2005809), a Korea Basic Science Institute (National Research Facilities and Equipment Center) grant funded by the Ministry of Education (No. 2021R1A6C101B418). Correspondence and requests for materials should be addressed to DK (dohunkim@snu.ac.kr).

**Contributions**

D.K. and W.J. conceived the project. W.J. performed the measurements with J.K. W.J analyzed the data. J.K. and H.Jung fabricated the device. J.P., M.C., H.Jang, and S.S. built the experimental setup and configured the measurement software. V.U. synthesized and provided the GaAs heterostructure. All the authors contributed to the preparation of the manuscript.

# Supplementary Notes: True decoherence-free-subspace derived from a semiconductor double quantum dot Heisenberg spin-trimer

Wonjin Jang[1,2], Jehyun Kim[1,3], Jaemin Park[1], Min-Kyun Cho[1,4], Hyeongyu Jang[1], Sangwoo Sim[1], Hwanchul Jung[5], Vladimir Umansky[3], and Dohun Kim[1]*

[1]NextQuantum, Department of Physics and Astronomy, and Institute of Applied Physics, Seoul National University, Seoul 08826, Korea

[2]Institute of Physics and Center for Quantum Science and Engineering, École polytechnique fédérale de Lausanne (EPFL), Lausanne, 1015, Switzerland

[3]Braun Center for Submicron Research, Department of Condensed Matter Physics, Weizmann Institute of Science, Rehovot 76100, Israel

[4]Center for Superconducting Quantum Computing System, Korea Research Institute of Standards and Science, Daejeon 34113, Korea

[5]Department of Physics, Pusan National University, Busan 46241, Korea

*Corresponding author(s). E-mail(s): dohunkim@snu.ac.kr;

## Contents

# 1 Pulsed-gate spectroscopy of $\delta R_{ST}$

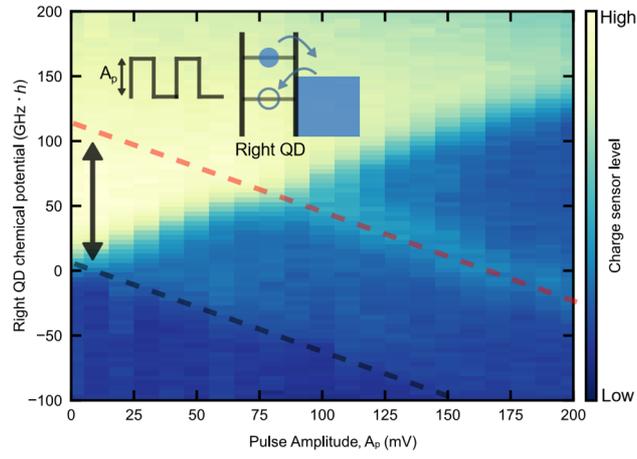

**Fig. S1 Pulsed-gate spectroscopy of $\delta R_{ST}$ [1].** The rf-SET sensor level is recorded as a function of the pulse amplitude $A_p$ and chemical potential of the right QD (see Fig. 2). A train of square pulses with an on (off) time of 100 ns (100 ns) is applied to $V_2$ (see Fig. 2a) to modulate the chemical potential of the right QD with respect to the Fermi level of the electron reservoir. The dashed black (red) line denotes the chemical potential where the ground (excited) orbital state and the Fermi level aligns under the pulse. $\Delta R_{ST} \sim 108$ GHz $\cdot h$ is extracted from the width indicated by the black arrow.

## 2 Toy model Hamiltonian of the Heisenberg spin trimer in a double quantum dot

Three electrons confined in a double quantum dot (DQD) with the corresponding charge configuration [2,1]-[1,2] can be described by the following Hamiltonian Eq. (S1) [2, 3]. The Hamiltonian is written with the ordered basis [$D_{S(+1/2)}[2,1]$, $D_{T(+1/2)}[2,1]$, $Q_{(+1/2)}[2,1]$, $D_{S(+1/2)}[1,2]$, $D_{T(+1/2)}[1,2]$, $Q_{(+1/2)}[1,2]$]. As in the main text, $D_S$, $D_T$, and $Q$ denotes the doublet-singlet, doublet-triplet, and quadruplet state with n (m) denoting the charge number in the left (right) quantum dot (QD) by the [n,m] notation. While the Hamiltonian is shown only for the $m_s = +1/2$ subspace for simplicity, we note that the description of the $m_s = -1/2$ subspace is identical.

$$H_{elec} = \begin{bmatrix} \eta\varepsilon/2 & 0 & 0 & t_{11} & t_{12} & 0 \\ 0 & \varepsilon/2 + \delta L_{ST} & 0 & t_{21} & t_{22} & 0 \\ 0 & 0 & \varepsilon/2 + \delta L_{ST} & 0 & 0 & t_{22} \\ t_{11} & t_{21} & 0 & -\varepsilon/2 & 0 & 0 \\ t_{12} & t_{22} & 0 & 0 & -\varepsilon/2 + \delta R_{ST} & 0 \\ 0 & 0 & t_{22} & 0 & 0 & -\varepsilon/2 + \delta R_{ST} \end{bmatrix} \quad (S1)$$

Here, $\varepsilon$ is the DQD detuning, $t_{ij}$ denotes the tunnel coupling strength between the $i^{th}$ and $j^{th}$ orbital level in different QDs, and $\delta L_{ST}$ ($\delta R_{ST}$) presents the orbital splitting in the left (right) QD. We also introduce $\eta$ to account for the different lever-arms of the ground and excited states influenced by Wigner molecularization [4].

The effect of the magnetic field difference between the QDs, $\Delta B_z = B_L^z - B_R^z$, can be represented by $H_{\Delta B}$ (Eq. (S2)) in the same ordered basis as in Eq. (S1). $B_L^z$ ($B_R^z$) is the magnetic field strength along the z-axis in the left (right) QD. For the $m_s = -1/2$ subspace, $H_{\Delta B}$ has the opposite sign.

$$H_{\Delta B} = \begin{bmatrix} B_L^z/2 & 0 & 0 & 0 & 0 & 0 \\ 0 & (4B_R^z - B_L^z)/6 & \sqrt{2}\Delta B_z/3 & 0 & 0 & 0 \\ 0 & \sqrt{2}\Delta B_z/3 & (2B_L^z + B_R^z)/6 & 0 & 0 & 0 \\ 0 & 0 & 0 & B_R^z/2 & 0 & 0 \\ 0 & 0 & 0 & 0 & (4B_L^z - B_R^z)/6 & -\sqrt{2}\Delta B_z/3 \\ 0 & 0 & 0 & 0 & -\sqrt{2}\Delta B_z/3 & (2B_R^z + B_L^z)/6 \end{bmatrix} \quad (S2)$$

For the numerical simulation of the eigenstates, we neglect the transverse magnetic field component, which further allows us to exclude $Q_{\pm 3/2}$ from the model due to the spin selection rule. Thereby, $H = H_{elec} \pm H_{\Delta B}$ successfully describes the dynamics in the $m_s = \pm 1/2$ subspace.

Throughout the work, the parameters presented in Table S1 are utilized for the numerical simulation of the eigenstates.

**Table S1** Hamiltonian Parameters

| Parameter | Value |
|---|---|
| $t_{11}$ | 2.2 (GHz · h) |
| $t_{12}$ | 2.6 (GHz · h) |
| $t_{21}$ | 2.6 (GHz · h) |
| $t_{22}$ | 1.7 (GHz · h) |
| $\delta L_{ST}$ | 1.323 (GHz · h) |
| $\delta R_{ST}$ | 108 (GHz · h) |
| $\eta$ | 1.004 |

# 3 Dynamic nuclear polarization sequence

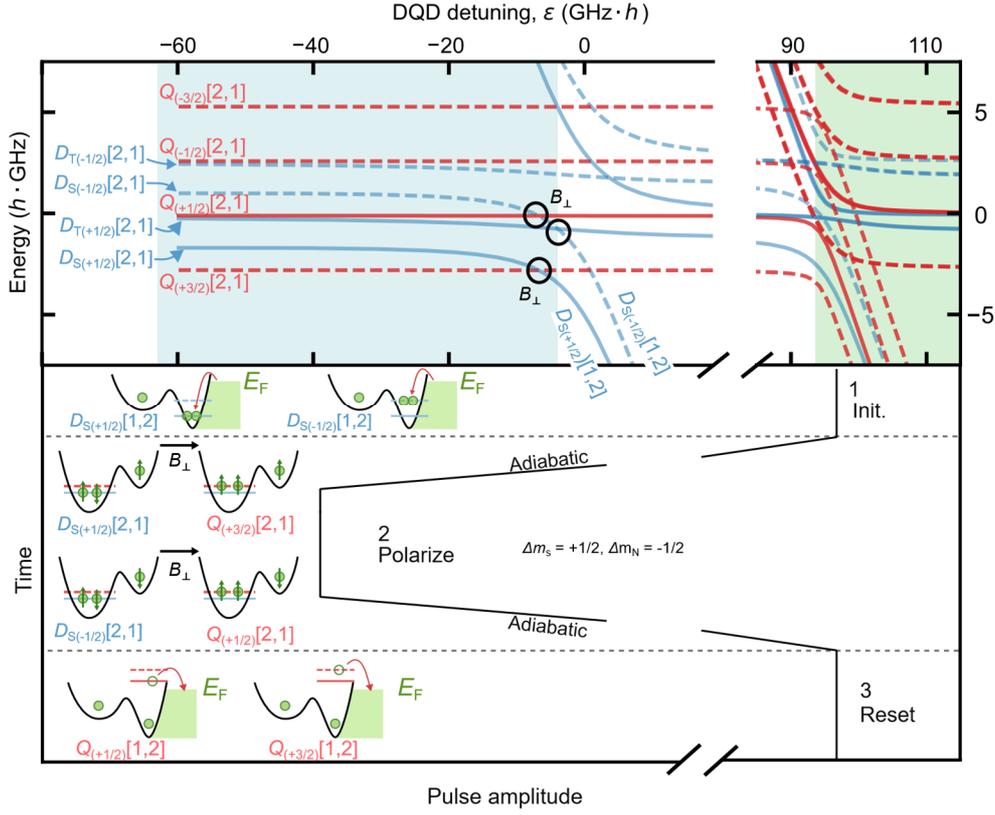

**Fig. S2 Dynamic nuclear polarization with the spin trimer**. Top panel: Eigenenergies as a function of the DQD detuning $\varepsilon$ (identical to Fig. 2d in the main text). Bottom panel: Pulsed-gate dynamic nuclear polarization scheme. See text for details.

In this section we detail the dynamic nuclear polarization (DNP) enabled by the pulsed-gate technique [5]. As shown in Supplementary Figure S2, a simple square pulse facilitates the DNP process. At the initialization stage, the spin state is prepared to either $D_{S(+1/2)}[1,2]$ or $D_{S(-1/2)}[1,2]$ by the energy-selective tunneling (EST) process which is then adiabatically brought to $D_{S(+1/2)}[2,1]$ or $D_{S(-1/2)}[2,1]$. During the polarization stage, if the DQD detuning reached by the pulse, or the pulse-tip $\varepsilon_p$, coincides with the $D_{S(+1/2)}[2,1]$-$Q_{(+3/2)}[2,1]$ ($D_{S(-1/2)}[2,1]$-$Q_{(+1/2)}[2,1]$ or $D_{S(-1/2)}[2,1]$-$D_{T(+1/2)}[2,1]$) degeneracy point, $D_{S(+1/2)}[2,1]$ ($D_{S(-1/2)}[2,1]$) can probabilistically evolve into $Q_{(+3/2)}[2,1]$ ($Q_{(+1/2)}[2,1]$ or $D_{T(+1/2)}[2,1]$) due to the finite $B_\perp$. This induces an electron spin flip with $\Delta m_s = +1$ which in turn flops a nuclear spin due to the conservation of angular momentum and increases the Zeeman splitting experienced by the electrons [5, 6]. As the final step, the states are adiabatically brought back to the initial detuning for resetting. Repeating such results in a sizable nuclear Overhauser field which also yields $\Delta B_z$ [5–7]. Because the transition from $D_{S(+1/2)}[2,1]$ to $Q_{(+3/2)}[2,1]$ involves a spin flip in the left QD, we expect the DNP to mainly generate the Overhauser field near this particular QD and yield $\Delta B_z > 0$.

# 4 Measurement of the $\Delta B_z$ fluctuation with a singlet-triplet qubit

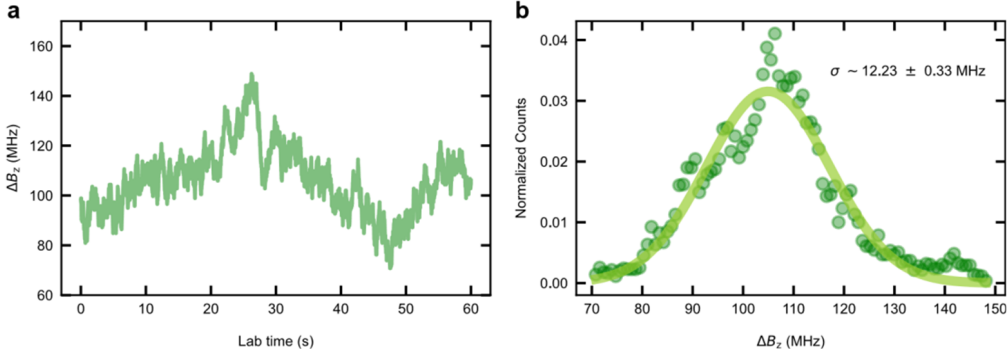

**Fig. S3 $\Delta B_z$ fluctuation measured with a singlet-triplet qubit. a,** An instance of the $\Delta B_z$ fluctuation acquired with the Bayesian estimation [8]. Each data point is estimated from 120 single-shot measurements taking 6 ms in total (~ 50 $\mu$s per measurement). **b,** Histogram of the trace shown in **a**. Numerical fit to a Gaussian yields the size of the $\Delta B_z$ fluctuation $\sigma \sim 12.23$ MHz. The same measurement is repeated multiple times to obtain the average fluctuation $\sigma_{\text{avg}} \sim 13.39$ MHz.

Within the same device, we also characterize the $\Delta B_z$ fluctuation by investigating the singlet-triplet (ST$_0$) qubit in the [1,1]-[0,2] charge configuration. A square pulse similar to that used in this work (Fig. 5) can also induce coherent Landau-Zener-Stuckelberg (LZS) oscillations of the ST$_0$ qubit, where the oscillation mainly reveals the relative phase evolution between the $|\uparrow,\downarrow\rangle$ and $|\downarrow,\uparrow\rangle$ states [6, 9]. Here, $|\uparrow,\downarrow\rangle$ ($|\downarrow,\uparrow\rangle$) refers to the spin state in the [1,1] charge configuration where the left QD has $\uparrow$ ($\downarrow$) and right QD has $\downarrow$ ($\uparrow$) spin. Noting that $|\uparrow,\downarrow\rangle$ and $|\downarrow,\uparrow\rangle$ have $\langle S_L^z - S_R^z\rangle$ of 1 and $-1$ respectively, the susceptibility of the $|\uparrow,\downarrow\rangle \leftrightarrow |\downarrow,\uparrow\rangle$ energy splitting to $\Delta B_z$, $\delta\langle S_L^z - S_R^z\rangle = 2$ holds. In other words, the ST$_0$ oscillation is not protected from the $\Delta B_z$ fluctuation and explicitly infers the size of $\Delta B_z$ fluctuation at the DQD site [6, 8].

Exploiting the Bayesian estimation, we also track the size of $\Delta B_z$ as a function of laboratory time [8, 10, 11]. Figure S3a presents an instance of the $\Delta B_z$ record, where each data point is estimated from 120 EST single-shot readouts [9], which takes 50 $\mu$s per measurement. Figure S3b shows the normalized histogram of the $\Delta B_z$ trace shown in Fig. S3a, from which we extract the size of the $\Delta B_z$ fluctuation of $\sigma \sim 12.23$ MHz. By repeating this measurement, we extract different values of the $\Delta B_z$ fluctuation as presented in Table S2. Discarding the largest and smallest values, we obtain the average $\sigma_{\text{avg}} \sim 13.39$ MHz.

| Run # | $\sigma$ (MHz) |
| --- | --- |
| Run#1 | 12.23 ± 0.33 |
| Run#2 | 15.07 ± 1.3 |
| Run#3 | 17.26 ± 2.6 |
| Run#4 | 14.54 ± 1.27 |
| Run#5 | 11.07 ± 0.13 |
| Run#6 | 17.71 ± 1.09 |
| Run#7 | 11.16 ± 0.31 |
| Run#8 | 12.89 ± 1.08 |
| Run#9 | 17.21 ± 1.46 |
| Run#10 | 12.95 ± 0.81 |
| Run#11 | 34.00 ± 39.35 |
| Run#12 | 15.94 ± 1.3 |
| Run#13 | 12.44 ± 1.03 |
| Run#14 | 6.46 ± 0.13 |
| Run#15 | 9.02 ± 0.3 |
| Run#16 | 11.05 ± 0.61 |
| Run#17 | 10.41 ± 0.51 |

**Table S2** Size of the $\Delta B_z$ fluctuations

# Supplementary References